\documentclass[preprint,aps]{revtex4}
\usepackage{epsfig}
\usepackage{amsmath}

\begin{document}

\title{Entanglement Switch for Dipole Arrays}

\author{Qi Wei and Sabre Kais\footnote{Corresponding Author: kais@purdue.edu}}
\address{Department of Chemistry and Birck Nanotechnology Center,
Purdue University, West Lafayette, IN 47907, USA}
\author{Yong P. Chen}
\address{Department of Physics  and Birck Nanotechnology Center,
Purdue University, West Lafayette, IN 47907, USA}

\begin{abstract}
We propose a new  entanglement switch of qubits consisting of
electric dipoles, oriented along or against an external electric
field and coupled by the electric dipole-dipole interaction. The
pairwise entanglement can be tuned and controlled by the ratio of
the Rabi frequency and the dipole-dipole coupling strength. Tuning
the entanglement can be achieved for one, two and three-dimensional
arrangements of the qubits. The feasibility of building such an
entanglement switch is also discussed.

\end{abstract}

\maketitle

\newpage

Entanglement is a quantum mechanical property that describes a
correlation between quantum mechanical systems. It has no classical
analog and has been lying in the heart of the foundation of quantum
mechanics. The desire to understand, tune and manipulate quantum
entanglement is of fundamental importance in the field of quantum
information and computation\cite{book,Amico,Vedral,kais}. Recently,
we studied a set of localized spins coupled through exchange
interaction and subject to an external magnetic filed
\cite{Osenda1,Osenda2,Osenda3,Osenda4}. We demonstrated for such a
class of one-dimensional magnetic systems, that entanglement can be
controlled and tuned by varying the anisotropy parameter in the
Hamiltonian and by introducing impurities into the
systems\cite{kais}. In this letter, we propose a new  entanglement
switch of qubits consisting of the electric dipole moment of
diatomic polar molecules, oriented along or against an external
electric field and coupled by the electric dipole-dipole
interaction.

Recent progress in  methods for producing, trapping and controlling
cold polar molecules make them an excellent candidate for quantum
computation\cite{Book2009,Herschbach,Fredrich,Demille}. Trapped
polar molecules was proposed as a novel physical realization of a
quantum computer by a number of authors
\cite{Demille,Gavin1,Gavin2,Micheli, Lukin, Jaksch,yelin,cha,kuz}.
In this proposal, the qubits are the molecular electric dipole
moments which can only orient along($|0\rangle$) or against
($|1\rangle$) the external electric fields. Each qubit is one polar
molecule with equal spacing along the axis. For such a system,
tuning and controlling the entanglement between the dipoles is of
great importance.

The Hamiltonian of N-trapped dipoles in an external electric field
reads\cite{chapter},
\begin{equation}
{\bf H}=\hbar\sum_{i=1}^{N}\omega_i \hat{S}_i^z +
\hbar\sum_{i\not=j}^{N}\Omega_{ij} \hat{S}_i^+\hat{S}_j^-,
\end{equation}
where $\hat S$'s are the dipole operators and related to Pauli
matrices, $\hat S_i^+$ and $\hat S_j^-$ represent dipole excitation
and de-excitation, respectively. $\omega_i$ is the transition
frequency of the dipole on site $i$, which is a function of dipole
moment and external electric fields at site $i$.
\begin{equation}
\hbar\omega_i = \left |\vec{d}\cdot\vec{E}\right |
\end{equation}
where $\vec{d}$ is the electric dipole moment which we assume the
same for each site. $\Omega_{ij}$ is the dipole-dipole coupling
constant between sites $i$ and $j$, which is determined by the
strength of dipole moment, the direction of the external electric
field and the lattice constant of the dipole array.
\begin{equation}
\hbar\Omega_{ij} =
\frac{|\vec{d}|^2(1-3cos^2\theta)}{|\vec{r}_{ij}|^3}
\end{equation}
where $\theta$ is the angle between $\vec{r}_{ij}$ and external
electric field.

We will expand the Hamiltonian in pure standard basis,
$\{|00\rangle,|01\rangle,|10\rangle,|11\rangle\}$,  and all the
eigenstates will be obtained  by diagonalizing the Hamiltonian
matrix. For example, for $N=2$, we obtained the following four
eigenvectors: $\Psi_1=|00\rangle, \;
|\Psi_2=\frac{1}{\sqrt{2}}(|01\rangle-|10\rangle), \;
|\Psi_3=\frac{1}{\sqrt{2}}(|01\rangle+|10\rangle),
\;\Psi_4=|11\rangle$ with corresponding eigenvalues:
$E_1=0,\;E_2=\omega-\Omega, \; E_3=\omega+\Omega, \; E_4=2 \omega$.
For the ratio $\omega/\Omega < 1$, $\Psi_2$ is fully entangled
ground state. However, for $\omega/\Omega > 1$ the non entangled
state $\Psi_1$ is the ground state. Thus, we have a curve crossing
at $\omega/\Omega=1$.

The concept of entanglement of formation is related to the amount of
entanglement needed to prepare the state $\rho$, where $\rho$ is the
density matrix. It was shown by Wootters \cite{Wootters} that
concurrence is a good measure of entanglement\cite{Hill}. The
concurrence C is given by \cite{Wootters}
\begin{equation}
C(\rho) = \max \{0,\lambda_1-\lambda_2-\lambda_3-\lambda_4 \}
\end{equation}
For a general state of two qubits, $\lambda_i$'s are the
eigenvalues, in decreasing order, of the Hermitian matrix $R \equiv
\sqrt{\sqrt{\rho}\tilde{\rho}\sqrt{\rho}}$ where $\rho$ is the
density matrix and $\tilde{\rho}$ is the spin-flipped state defined
as
\begin{equation}
\tilde{\rho} =
(\sigma_y\otimes\sigma_y)\rho^*(\sigma_y\otimes\sigma_y)
\end{equation}
 where the $\rho^*$ is the complex
conjugate of $\rho$ and is taken in the standard basis, which for a
pair of 2 level particles is
$\{|00\rangle,|01\rangle,|10\rangle,|11\rangle\}$.

In order to calculate thermal entanglement, we need the temperature
dependent density matrix and the density matrix for a system in
equilibrium at a temperature T reads: $\rho = e^{-\beta \hat H /Z}$
with $\beta = \frac{1}{kT}$ and $Z$ is the partition function, $Z =
\text{\bf {Tr}} (e^{-\beta \hat H})$. In this case, the partition
function is
\begin{equation}
Z(T) = \sum_{i}g_ie^{-\beta \lambda_i}
\end{equation}
where $\lambda_i$ is the $i$th eigenvalue and $g_i$ is the
degeneracy. And the corresponding density matrix can be written:
\begin{equation}
\rho(T) = \frac{1}{Z}\sum_{i}^{N}e^{-\beta
\lambda_i}|\Phi_i\rangle\langle\Phi_i|
\end{equation}
where $|\Phi_i\rangle$ is the $i$th eigenfunction. For pairwise
thermal entanglement, we can get reduced density matrix as a
function of temperature in the same way, which leads to temperature
dependent entanglement.

In Figure 1 we show the tuning of the pairwise entanglement,
measured by concurrence, of one-dimensional arrangements of the
dipoles as one varies the ratio $\omega/\Omega$ at different
temperatures for $N=9$ dipoles. Here we took all transition
frequencies to be the same, $\omega_i=\omega$ and the nearest
neighbor dipole-dipole interaction to be the same,
$\Omega_{i,i+1}=\Omega_{i,i-1}=\Omega$. All other dipole-dipole
coupling constants $\Omega_{i,j\neq i\pm 1}$ can be expressed in
terms of $\Omega$.  Thus, we have two parameters to vary, the ratio
$\omega/\Omega$ and temperature $kT$. At $kT \sim 0 $ one has a
constant entanglement over  a long ratio $\omega/\Omega$ and sharp
transitions or jumps to lower values at other values of
$\omega/\Omega$. It is worth mentioning that for $\omega/\Omega <
0.634$, entanglement is only between dipole 1 and 2. For $
0.634<\omega/\Omega < 1.14$ dipole one becomes entangled also with
dipole 3 and with other dipoles until we reach $ \omega/\Omega \sim
1.74$, above this value the concurrence is zero between all sites.
As one increase the temperature, the curve become smoother as mixing
occur with higher states. Calculations for $N=3,4,...,8$ gave
similar behavior of tuning and controlling entanglement as for the
case $N=9$. To show how the populations changes at the transition
point, we present in Figure 2 the coefficients of the wave function
for $N=4$ before and after the transition point $\omega/\Omega
=0.64$. When $\omega/\Omega < 0.64$, the ground state wave function
is
$|\Psi\rangle=0.19|\downarrow\downarrow\uparrow\uparrow\rangle-0.51|\downarrow\uparrow\rangle+0.45|\downarrow\uparrow\uparrow\downarrow\rangle+0.45|\uparrow\downarrow\downarrow\uparrow\rangle-0.51|\uparrow\downarrow\uparrow\downarrow\rangle+0.19|\uparrow\uparrow\downarrow\downarrow\rangle$,
however, when $\omega/\Omega > 0.64$, the ground state wave function
becomes
$|\Psi\rangle=0.36|\downarrow\downarrow\downarrow\uparrow\rangle-0.61|\downarrow\downarrow\uparrow\downarrow\rangle+0.61|\downarrow\uparrow\downarrow\downarrow\rangle-0.36|\uparrow\downarrow\downarrow\downarrow\rangle$.
Figures 3 and 4 show a similar phenomena for two and
three-dimensional arrangements of the dipoles.  The pairwise
entanglement decreases as
one increases the dimensionality of the system and the temperature.  \\

There have been rapid advances made recently in cooling, trapping
and manipulating atomic (Rydberg)\cite{wisc,amt,pfau} and molecular
dipoles\cite{ni,lics,carr,Herschbach,Fredrich}. For example, a wide
variety of ground state polar molecules with large electric dipole
moments (several Debyes, where 1 Debye $\sim$ 3.3 $\times$
$10^{-30}$\text{ Cm}) \text { have been cooled to ultracold} ($\sim$
mK or below) regime, some even near quantum degeneracy \cite{carr}.
These developments provide exciting opportunities to experimentally
realize the entanglement switch described above, which requires
tuning $\omega/\Omega$ around 1, and cooling the dipoles to
temperatures ($T$) corresponding to a fraction of the dipole-dipole
interaction energy ($\Omega$). For a dipole moment ($p$) of few
Debyes, and a typical experimental electric field ($\sim\;$10$^5$
V/m, required for the dipole moment to actually manifest itself
\cite{bohn,Demille}), dipole-dipole separation ($d$) on the order of
10 nm is required for $\omega/\Omega$ $\sim$ 1. Such $d$ is much
shorter than what can be achieved in typical optical lattices (as
envisaged in \cite{Demille}, which corresponds to the regime of
$\omega/\Omega$ $\gg$ 1), but can be realized with arrays of
nanoscale plasmon-enhanced electric/electro-optical traps recently
proposed\cite{chang,murphy} (where extremely tight, few-nanometer
confinement and trap frequencies exceeding 100MHz are shown to be
possible). At such short $d$, the dipole-dipole interaction
$\Omega=\frac{p^2}{4\pi \epsilon_0 d^3}$ is on the order of 0.1 K.
Cooling to a small fraction of such a temperature scale is easily
within the current experimental technology. Furthermore,
$\omega/\Omega$ can be tuned experimentally by the electrical field,
and/or by $d$ (eg. by varying the microtrap configuration).

In summary, we presented  a new way  to construct an entanglement
switch in an optical lattice and discussed  the experimental
feasibility of building such switch. The realization of such a
scheme will have a profound impact on the implementation of quantum
gates  in quantum computing with trapped polar molecules. The
similar idea could in principle also be applied to other systems
with electric dipoles, such as quantum dots, excitons in
nanostructurs.

\section{Acknowledgment}
We would like to thank the ARO for financial support.
YPC acknowledges partial support from National Science Foundation.

\newpage

\begin{figure}
\begin{center}
\includegraphics[width=0.50\textwidth]{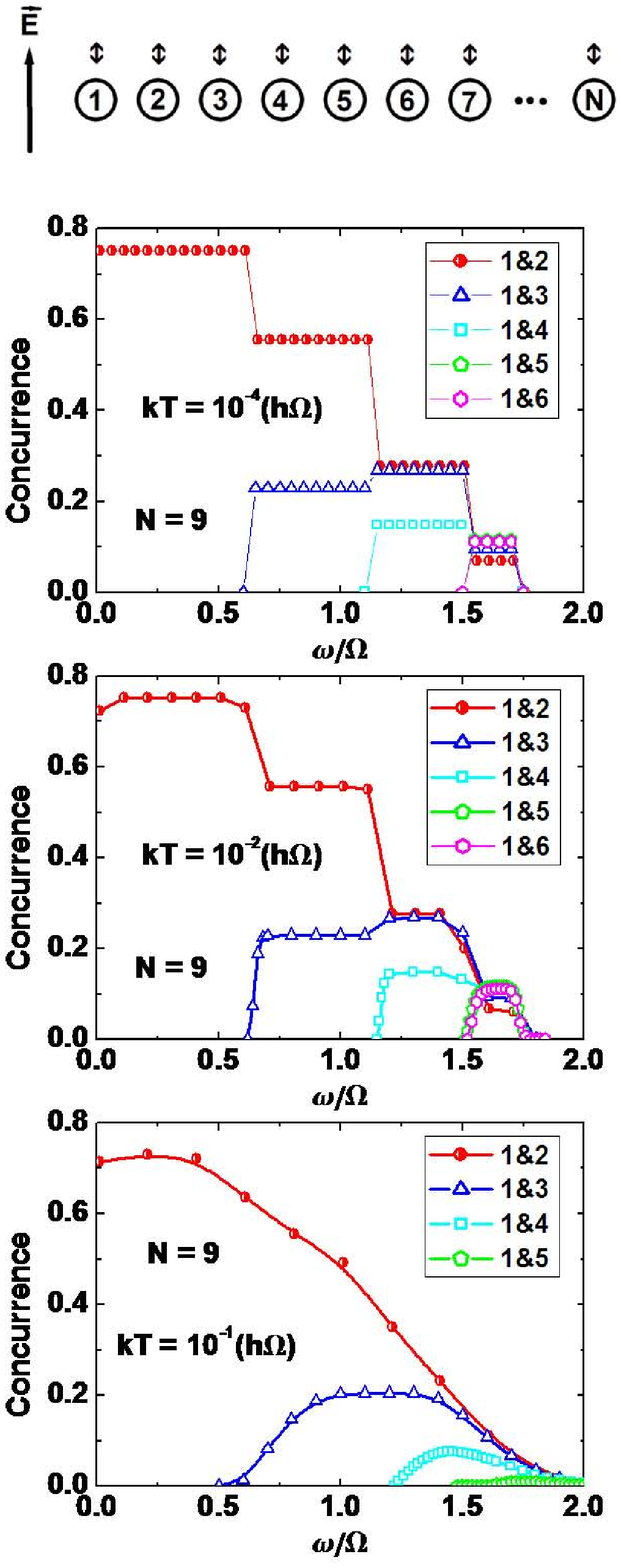}
\end{center}
\caption{\it Pairwise concurrence (Eq. 4) of one-dimensional
arrangements of the dipoles as one varies the ratio $\omega/\Omega$
at $kT=10^{-4},\;10^{-2}\text{ and }10^{-1}\;\hbar\Omega$
respectively for $N=9$ dipoles.}
\end{figure}

\newpage

\begin{figure}
\begin{center}
\includegraphics[width=0.60\textwidth]{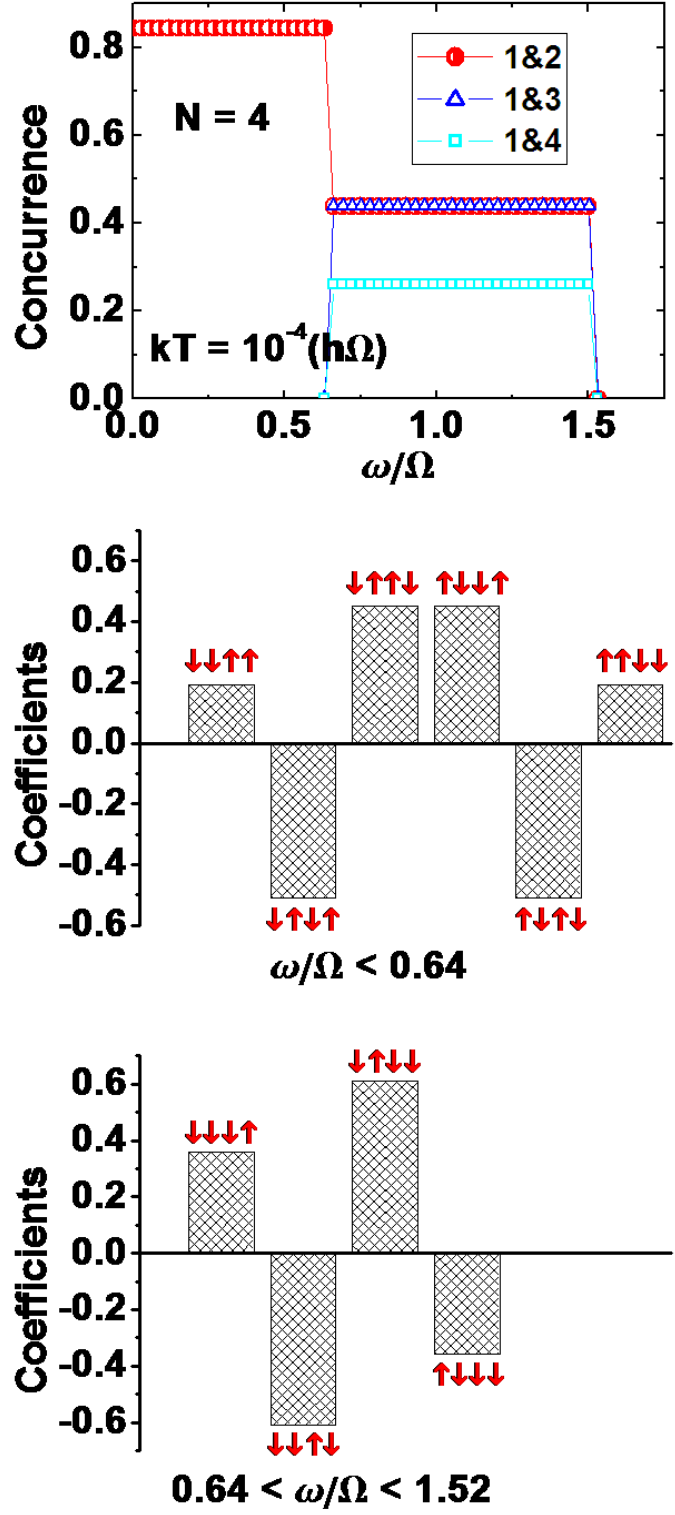}
\end{center}
\caption{\it Pairwise concurrence (Eq. 4) and coefficients of the
wave function of one-dimensional arrangements of the dipoles as one
varies the ratio $\omega/\Omega$ at $kT=10^{-4}\;\hbar\Omega$ for
$N=4$ dipoles.}
\end{figure}

\newpage

\begin{figure}
\begin{center}
\includegraphics[width=0.90\textwidth]{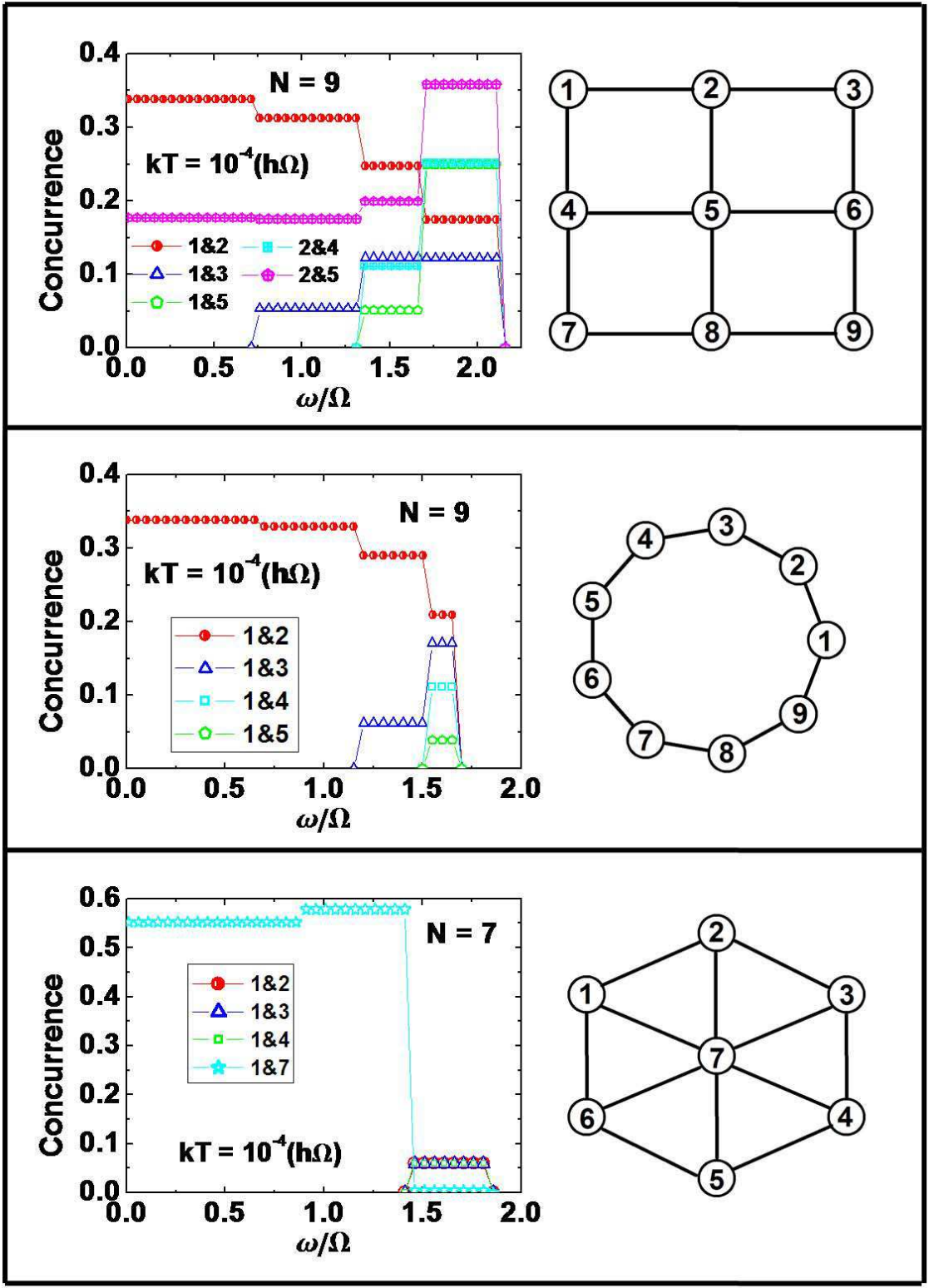}
\end{center}
\caption{\it Pairwise concurrence (Eq. 4) for different
two-dimensional arrangements of the dipoles as one varies the ratio
$\omega/\Omega$ at $kT=10^{-4}\;\hbar\Omega$. The external electric
field is perpendicular to the 2-d plane.}
\end{figure}

\newpage

\begin{figure}
\begin{center}
\includegraphics[width=0.95\textwidth]{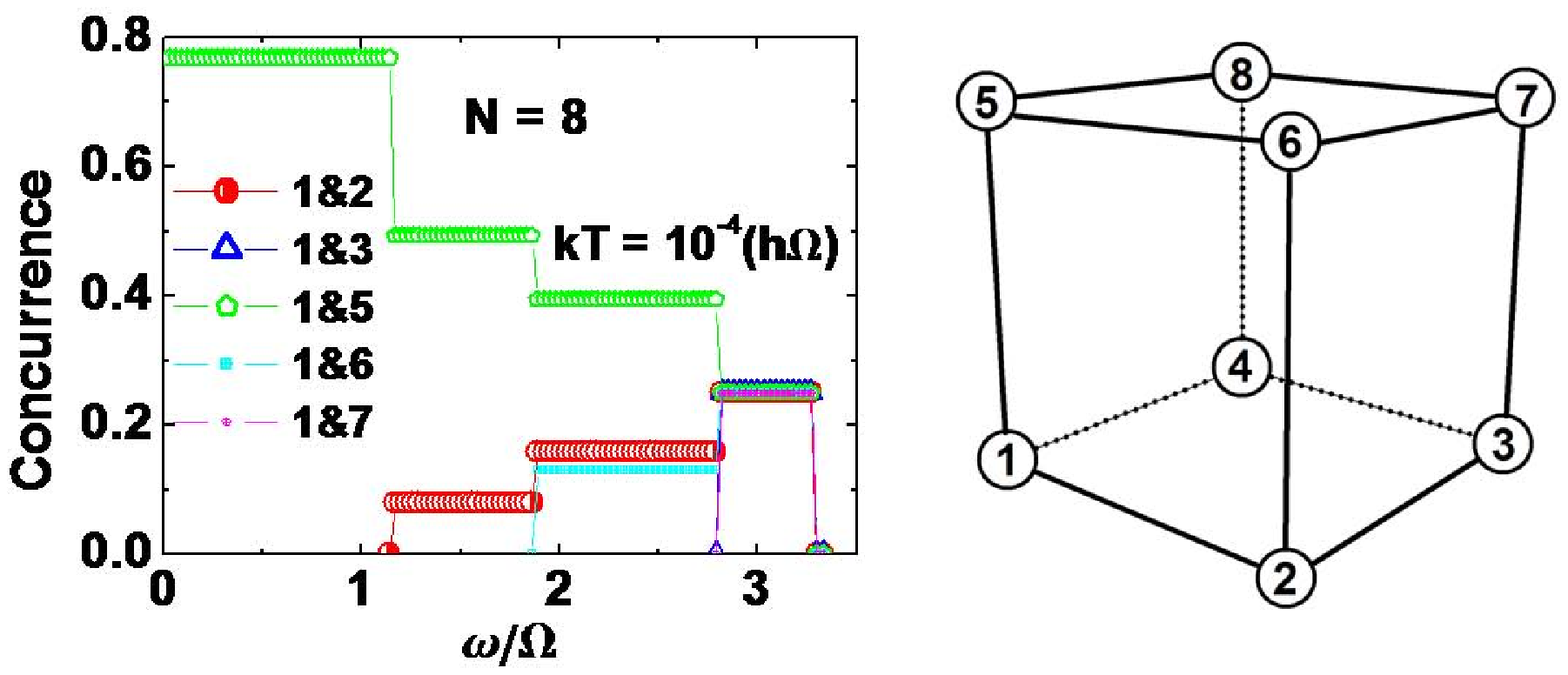}
\end{center}
\caption{\it Pairwise concurrence (Eq. 4) of three-dimensional
simple cubic arrangements of the dipoles as one varies the ratio
$\omega/\Omega$ at $kT=10^{-4}\;\hbar\Omega$. Here $\Omega$ is the
dipole-dipole coupling constant between dipole 1 and dipole 2,
$\Omega\;=\;\Omega_{12}$.}
\end{figure}


\begin{thebibliography}{33}

\bibitem{book} M. A. Nielsen, I. L. Chuang,   Quantum Computation
and Quantum Information, (Cambridge: University Press, 2000).

\bibitem{Amico} L. Amico, R. Fazio , A. Osterloh , and V. Vedral ,
Rev. Mod. Phys. {\bf 80}, 517 (2008).

\bibitem{Vedral} V. Vedral,
Nature, {\bf 453}, 1004 (2008).

\bibitem{kais} S. Kais,
Adv.  Chem.  Phys. {\bf  134}, 493 (2007).

\bibitem{Osenda1} O. Osenda, Z. Huang and S. Kais,
Phys. Rev. A {\bf 67}, 062321-4 (2003).

\bibitem{Osenda2} Z. Huang, O. Osenda and S. Kais,
Phys. Lett. A {\bf 322}, 137 (2004).

\bibitem{Osenda3} Z. Huang and S. Kais, In. J. Quant. Information {\bf 3}, 83 (2005).

\bibitem{Osenda4} Z. Huang and S. Kais, Phys, Rev. A {\bf 73}, 022339 (2006).

\bibitem{Book2009} R. Krems, W.C. Stwalley, and B. Friedrich,
"Cold molecules: theory, experiment, applications", (Taylor and
Francis (2009)).

\bibitem{Herschbach} D. Herschbach,
Molecular collisions, from warm to ultracold. \textit{Faraday Discussions} \textbf{142}, 9-23 (2009)

\bibitem{Fredrich} B. Friedrich and J.M. Doyle, ChemPhysChem \textbf{10}, 604 (2009).

\bibitem{Demille} D. Demille,  \textit{Phys. Rev. Lett.} {\bf 88}, 067901 (2002).

\bibitem{Gavin1}G.K. Bernnen, I.H. Deutsch, P.S. Jessen,
Phys. Rev. A {\bf 61}, 062309-1 (2000).

\bibitem{Gavin2}G. K. Brennen, C. M. Caves, P. S. Jessen
and I.H. Deutsch,  Phys. Rev. Lett. {\bf 82}, 1060 (1999).

\bibitem{Micheli} A. Micheli, G. K. Brennen and P. Zoller,  \textit{Nature Phys.} {\bf 2}, 341-347 (2006).

\bibitem{Lukin} M. D. Lukin \textit{et al},
 \textit{Phys. Rev. Lett.} {\bf 87}, 037901 (2001).

\bibitem{Jaksch} D. Jaksch, J. I. Cirac, P. Zoller,
 \textit{Phys. Rev. Lett.} {\bf 85}, 2208 (2000).

\bibitem{chapter} Z. Ficek and S. Swain,
 {\it Quantum Interference and Coherence Theory and Experiment},
(Springer, NY 2004).

\bibitem{Wootters} W. K. Wootters,
\textit{Phys. Rev. Lett.} {\bf 80}, 2245 (1998).

\bibitem{Hill} S. Hill  and W. K. Wootters,
\textit{Phys. Rev. Lett.} {\bf 78}, 5022 (1997).

\bibitem{yelin} S. F. Yelin, K. Kirby, and R. Cote,
\textit{Phys. Rev. A} 74, 050301(R) (2006).

\bibitem{cha} E.  Charron, P. Milman, A. Keller and O. Atabek,
\textit{Phys. Rev. A} {\bf 75}, 033414 (2007).

\bibitem{kuz} E. Kuznetsova, R. Cote, K. Kirby and S. F. Yelin,
\textit{Phys. Rev. A} {\bf 78}, 012313 (2008).

\bibitem{wisc} E. Urban, T. A. Johnson, T. Henage, L. Isenhower, D. D. Yavuz, T. G. Walker and M. Saffman,
\textit{Nature Phys.} {\bf 5}, 110 (2009).

\bibitem{amt}  T. Amthor, C. Giese, C. S. Hofmann and M. Weidemuller,
\textit{arXiv:0909.0837} (2009).

\bibitem{pfau} R. Heidemann, U. Raitzsch, V. Bendkowsky, B. Butscher, R. Low and T. Pfau,
\textit{Phys. Rev. Lett.} {\bf 100}, 033601 (2008).

\bibitem{ni} K. K. Ni \textit{et al.},
\textit{Science} {\bf 322}, 231 (2008).

\bibitem{lics} J. Deiglmayr \textit{et al.},
\textit{Phys. Rev. Lett.} {\bf 101}, 133004 (2008).

\bibitem{carr} L. D. Carr, D. DeMille, R. V. Krems and J. Ye,
\textit{New Journal of Physics} {\bf 11}, 055049 (Focus Issue).
(2009)

\bibitem{aym} M. Aymar and O. Dulieua,
\textit{J. Chem. Phys.} {\bf 122}, 204302 (2005).

\bibitem{bohn} J. L. Bohn,
\textit{arXiv:0901.0276} (2009).

\bibitem{chang} D. E. Chang \textit{et al.},
\textit{Phys. Rev. Lett.} \textbf{103}, 123004 (2009)

\bibitem{murphy} B. Murphy and L. V. Hau,
\textit{Phys. Rev. Lett.} \textbf{102}, 033003 (2009).


\end{thebibliography}
\end{document}